\documentstyle[preprint,aps,epsf,floats]{revtex} 

\begin{document}

\def\ltap{\ \raise.3ex\hbox{$<$\kern-.75em\lower1ex\hbox{$\sim$}}\ }

\preprint{\tighten \vbox{\hbox{UCSD/PTH 99-20} \hbox{hep-ph/99mmnnn} }}

\title{Baryon-Pion Scattering in the $1/N_c$ Expansion: Tree Diagram Cancellations}

\author{Ruben Flores-Mendieta, Christoph P. Hofmann and Elizabeth Jenkins}

\address{Department of Physics, University of California at San Diego, La Jolla, CA 92093}

\maketitle

{\tighten
\begin{abstract}%
Tree amplitudes for baryon-pion scattering are studied in the $1/N_c$ expansion.  Generalized
large-$N_c$ consistency conditions are obtained to all orders in
baryon mass splittings.  For baryons with spin $J \sim {\cal O}(1)$,
the leading order in $N_c$ tree amplitudes can be evaluated keeping only terms up to a given
finite order in baryon mass splittings. 
\end{abstract}
}

\newpage

\section{Introduction}

The study of baryons in the $1/N_c$ expansion has led to significant theoretical progress
in understanding the spin-flavor structure of QCD baryons 
(See Ref.~\cite{jannrev} for a recent review).  In the large-$N_c$ limit, there exists a
spin-flavor symmetry for baryons~\cite{dm93,j93,djm94}, and  
baryons form irreducible representations of the
spin-flavor symmetry.  For $N_c$ large but finite, these irreducible representations contain baryons
with spins ranging from $J \sim {\cal O}(1)$ to $J \sim {\cal O}(N_c)$.  Important symmetry relations
can be derived for baryons with spin $J \sim {\cal O}(1)$.

Dashen, Jenkins and Manohar~\cite{djm94} showed that large $N_c$ power counting rules for
multipion--baryon-baryon scattering amplitudes yield important constraints on baryon axial vector
couplings and masses, as well as other static properties.  
Consistency of the large $N_c$ limit requires exact
cancellations amongst the tree diagram amplitudes at leading orders in $N_c$.  
These cancellations have been discussed explicitly in
Refs.~\cite{dm93,djm94} for 
the simplest cases of baryon-pion scattering to a baryon plus one or two pions.
The non-degeneracy of the baryon states in a given spin-flavor representation for finite $N_c$ 
results in additional consistency conditions involving
the baryon mass operator~\cite{j93}.  Interestingly, for baryons with two flavors of light quarks
$N_F =2$ and spin $J \sim {\cal O}(1)$, 
the leading contribution to the tree amplitude
for baryon-pion scattering to a baryon and a single pion comes from terms which are first order in the
baryon mass splittings~\cite{djm94}.

In this paper, we generalize the prior analysis of baryon-pion scattering amplitudes by Dashen,
Jenkins and Manohar~\cite{djm94}.  We obtain new large-$N_c$ consistency conditions for baryon axial
vector couplings by studying tree
scattering amplitudes to all orders in baryon mass splittings.  We also obtain additional large-$N_c$
consistency conditions for baryon vector current couplings.  Lam and Liu~\cite{lamliu} 
previously showed that the requisite cancellations of Ref.~\cite{djm94} occur in the degeneracy limit 
for tree scattering
amplitudes containing an arbitrary number of single pion--baryon-baryon vertices. 
The analysis of this work
extends this result to all orders in baryon mass splittings.
As we have remarked, terms involving baryon mass splittings contribute to baryon-pion
scattering amplitudes at leading order in $N_c$, so the demonstration of consistency of the large-$N_c$
limit for baryon-meson tree amplitudes is incomplete without the inclusion of baryon mass splittings.  

The organization of this paper is as follows.  In Sec.~II, we give a brief description of 
large-$N_c$ baryon spin-flavor symmetry and review baryon chiral perturbation theory in
the $1/N_c$ expansion.      
In Sec.~III, we study baryon-pion scattering amplitudes, and derive large-$N_c$
consistency conditions to all orders in baryon mass splittings.  We then compute the 
leading in $N_c$ tree amplitude explicitly for a few specific examples, 
and comment upon the general case.  
We conclude in Sec.~IV.

\section{Baryon Chiral Perturbation Theory in the $1/N_c$ Expansion}

The lowest-lying baryons for large $N_c$ are given by
the completely symmetric spin-flavor representation of $N_c$ quarks. 
Under $SU(2) \otimes SU(N_F)$, this $SU(2N_F)$ representation decomposes into a tower of
baryon flavor representations with spins $J= \frac 1 2, \frac 3 2, \cdots, \frac {N_c} 2$.  For two
flavors of light quarks $u$ and $d$, the baryon tower consists of (spin, isospin) representations with
$I=J$, while for three flavors of quarks, the baryon flavor representations are considerably more
complicated~\cite{djm94,djm95}.  It is advantageous to concentrate on the baryon
operators, rather than on the states, since baryon operators have a simple expansion in $1/N_c$ for
arbitrary $N_c$.

The general form of the $1/N_c$ expansion of a QCD $m$-body quark operator acting on a single 
baryon state is given by
\begin{eqnarray}
{\cal O}^{\rm m-body}_{\rm QCD} = N_c^m \sum_n c_n {1 \over N_c^n}
{\cal O}_n,
\end{eqnarray}
where the ${\cal O}_n$, $0\le n \le N_c$, 
are a complete set of linearly independent operator products 
which are of $n^{\rm th}$ order
in the baryon spin-flavor generators, and the $c_n(1/N_c)$ are arbitrary
unknown coefficients with an expansion in $1/N_c$ beginning at order unity.
The large-$N_c$ spin-flavor symmetry for baryons is generated by the baryon spin, flavor and
spin-flavor operators $J^i$, $T^a$ and $G^{ia}$ which can be written for large, but finite, $N_c$
as one-body quark operators
acting on the $N_c$-quark baryon states:
\begin{eqnarray}\label{generators}
J^i &=& q^\dagger \left( {\sigma^i \over 2} \otimes I \right) q, \nonumber\\
T^a &=& q^\dagger \left( I \otimes {\lambda^a \over 2} \right) q, \\
G^{ia} &=& q^\dagger \left( {\sigma^i \over 2} \otimes {\lambda^a \over 2} \right) q, \nonumber
\end{eqnarray}
where without loss of generality the baryon matrix elements of these operators can be taken
as the values in the non-relativistic quark model.\footnote{This convention is referred to as the quark
representation in the literature, see Ref.~\cite{djm95}.}
The baryon spin-flavor operators satisfy the $SU(2N_F)$ algebra given in Table~I.   
The operator basis ${\cal O}_n$ for any QCD operator transforming according to a given spin $\otimes$
flavor representation can be determined.  Examples of $1/N_c$ expansions
for baryon operators include the $1/N_c$ expansion of the 
baryon mass operator~\cite{j93,lmr,cgo,djm94,djm95}  
\begin{eqnarray}\label{eq:mass}
{\cal M }&=&  \sum_{n=0}^{(N_c-1)/2} m_{2n} {1
\over N_c^{2n-1}} \left(J^2\right)^n \nonumber\\
&=& m_0 N_c \openone + m_2 {1 \over N_c} J^2 + m_4 {1 \over N_c^3} J^4 + \cdots,
\end{eqnarray}
and the $1/N_c$ expansion of the baryon axial vector current~\cite{djm95}
\begin{eqnarray}\label{eq:axcurr}
A^{ia} = a_1 G^{ia} + \sum_{n=2,3}^{N_c} b_n \frac{1}{N_c^{n-1}} {\cal 
D}_n^{ia} + \sum_{n=3,5}^{N_c} c_n \frac{1}{N_c^{n-1}} {\cal O}_n^{ia} \, . 
\end{eqnarray}
In Eq.~(\ref{eq:axcurr}),
the ${\cal D}_n^{ia}$ are diagonal operators with nonzero matrix
elements only between baryon states with the same spin, whereas ${\cal O}_n^{ia}$
are purely off-diagonal operators with nonzero matrix elements only
between baryon states of different spin. The explicit forms for these operators
can be found in Ref.~\cite{djm95}. At the physical value $N_c = 3$, Eq.~(\ref{eq:mass})
reduces to 
\begin{eqnarray}
{\cal M }= m_0 N_c \openone + m_2 {1 \over N_c} J^2 ,
\end{eqnarray}
and Eq.~(\ref{eq:axcurr}) reduces to
\begin{eqnarray}
A^{ia} = a_1 G^{ia} + b_2 \frac{1}{N_c} J^i T^a + b_3 \frac{1}{N_c^2}
{\cal D}_3^{ia} + c_3 \frac{1}{N_c^2} {\cal O}_3^{ia} \, ,
\end{eqnarray}
where
\begin{eqnarray}
{\cal D}_3^{ia} & = & \{ J^i, \{J^j, G^{ja} \} \} \, , \\
{\cal O}_3^{ia} & = & \{ J^2, G^{ia} \} - \frac12 \{ J^i, \{J^j, G^{ja} \}
\} \, .
\end{eqnarray}

The $1/N_c$ chiral Lagrangian describing the interactions of soft pions
is formulated in terms of the field
\begin{eqnarray}
\xi(x) = e^{ i \Pi(x) / f},
\end{eqnarray}
where $\Pi(x)$ is the nonet of Goldstone boson fields
\begin{eqnarray}
\Pi (x)={{\pi^a(x) \lambda^a} \over 2} + {{\eta^\prime(x) I} \over \sqrt{6}}
\end{eqnarray}
and $f \sim {\cal O}(\sqrt{N_c})$ is the pion decay constant.   
(The $\eta^\prime$ field is a Goldstone boson in the large-$N_c$ limit because $U(1)_A$ is
a symmetry that is broken only at order $1/N_c$ by the axial anomaly~\cite{thooft76}.)
The $1/N_c$ chiral Lagrangian for matter fields depends on the $\xi$ field through the
vector and axial vector currents
\begin{eqnarray}
{\cal V}^\mu &=& {1 \over 2}
\left( \xi \partial^\mu \xi^\dagger + \xi^\dagger \partial^\mu \xi \right),
\nonumber\\ 
{\cal A}^\mu &=& {i \over 2} 
\left( \xi \partial^\mu \xi^\dagger - \xi^\dagger \partial^\mu \xi \right) \, . 
\end{eqnarray}
The Goldstone boson nonet vector and axial vector currents appearing 
in the $1/N_c$ baryon chiral Lagrangian are given by
\begin{eqnarray}
{\rm Tr} \left( {\cal V}^0 \lambda^a \right) &=&  {\rm Tr} \left(
\left({1 \over {2f^2}}\left[ \Pi, \partial^0 \Pi \right] 
-{1 \over{4!}}{1 \over f^4} \left[ \Pi, \left[ \Pi,\left[ \Pi, \partial^0 \Pi \right]\right]\right] 
+ \cdots \right)
\lambda^a \right)
= {i \over {2f^2}}f^{abc} \pi^b \partial^0
\pi^c + \cdots, \nonumber\\
{\rm Tr} \left( {\cal V}^0  {{2I} \over \sqrt{6}} \right) &=& {\rm Tr} \left(
\left({1 \over {2f^2}}\left[ \Pi, \partial^0 \Pi \right] 
-{1 \over{4!}}{1 \over f^4} \left[ \Pi, \left[ \Pi,\left[ \Pi, \partial^0 \Pi \right]\right]\right] 
+ \cdots \right)
{{2I} \over \sqrt{6}} \right)= 0,
\end{eqnarray}
and
\begin{eqnarray}
{\rm Tr} \left( {\cal A}^i \lambda^a \right) &=&  {\rm Tr} \left(
\left({1 \over f}\nabla^i \Pi  - {1 \over {3!}} {1 \over f^3} \left[ \Pi, \left[ \Pi, \nabla^i \Pi
\right]\right] + \cdots \right)
\ \lambda^a \right)
= {1 \over f} \nabla^i \pi^a + {1 \over {3!f^3}}f^{abe}f^{cde} \pi^b \pi^c \nabla^i \pi^d 
+\cdots, \nonumber\\
{\rm Tr} \left( {\cal A}^i {{2I} \over \sqrt{6}} \right) &=& {\rm Tr} \left(
\left({1 \over f} \nabla^i \Pi - {1 \over {3!}} {1 \over f^3} \left[ \Pi, \left[ \Pi, \nabla^i \Pi
\right]\right] + \cdots \right)
\ {{2I} \over \sqrt{6}} \right)= {1 \over f} \nabla^i \eta^\prime,
\end{eqnarray}
respectively.  Notice that the $SU(3)$ singlet portion of the pion vector current vanishes identically,
and that the singlet portion of the pion axial vector current is proportional to a single 
derivatively coupled $\eta^\prime$. 

The $1/N_c$ chiral Lagrangian for baryons 
in the baryon rest frame is given by~\cite{j96}
\begin{eqnarray}\label{blag}
{\cal L}_{\rm baryon} = i{\cal D}^0 - {\cal M}_{\rm hyperfine} + 
{\rm Tr}\left( {\cal A}^i \lambda^A \right) A^{iA} + 
{1 \over N_c} {\rm Tr}\left({\cal A}^i {{2I} \over \sqrt{6}} \right) A^{i} + \cdots\, ,
\end{eqnarray}
where the ellipsis represents terms of higher order in the derivative and $1/N_c$ expansions
as well as terms involving explicit chiral symmetry breaking by the quark mass matrix.
In Eq.~(\ref{blag}), the covariant derivative is equal to
\begin{eqnarray}
{\cal D}^0 = \partial^0 \openone + {\rm Tr}\left( {\cal V}^0 \lambda^A \right) T^A 
= \partial^0 \openone + {\rm Tr}\left( {\cal V}^0 \lambda^a \right) T^a,
\end{eqnarray}
where $A=1,\cdots,9$, and $\lambda^9 \equiv 2I/\sqrt{6}$, and
the summation over the index $A$ in the covariant derivative reduces to a summation over $a=1,\cdots,8$
because the ninth component of the pion vector current vanishes identically.
The leading ${\cal O}(N_c)$ singlet portion of the baryon mass has been removed
from the $1/N_c$ baryon chiral Lagrangian by a phase redefinition of the baryon field as in 
heavy baryon chiral perturbation theory~\cite{jm255,jm259},
so the $1/N_c$ baryon chiral Lagrangian
depends only on baryon mass splittings.  The baryon hyperfine mass operator is given by
\begin{eqnarray}
{\cal M}_{\rm hyperfine} = m_2 {1 \over N_c} J^2 + m_4 {1 \over N_c^3} J^4 + \cdots \, .
\end{eqnarray}
The last two terms in Eq.~(\ref{blag}) describe the axial couplings of a baryon to pions.
The baryon axial vector current $A^{i9}\equiv A^i$ is defined in terms of 
Eq.~(\ref{eq:axcurr}) 
and the baryon
one-body operators 
\begin{eqnarray}
G^{i9} &=& {1 \over \sqrt{6}}J^i, \nonumber\\
T^9 &=& {1 \over \sqrt{6}} N_c \openone . 
\end{eqnarray}
Nonet flavor symmetry of the pion--baryon-baryon axial vector couplings is broken 
explicitly by the last term in Eq.~(\ref{blag}), which gives
a nonet symmetry-breaking contribution to the singlet baryon axial vector current 
$A^i$ at relative order $1/N_c$.
The baryon chiral Lagrangian has been written in the rest frame of the baryon
for notational simplicity; it is straightforward to rewrite the Lagrangian in
an arbitrary Lorentz frame in which the baryon travels with a fixed four-velocity $v^\mu$.

$SU(3)$-breaking baryon mass splittings arise from higher order terms in the chiral
Lagrangian containing insertions of the quark mass matrix.  The terms in the $1/N_c$ baryon chiral
Lagrangian containing one power of ${\cal M}_q = {\rm diag}(m_u,m_d,m_s)$ are given by
\begin{eqnarray}\label{lagmq}
{\cal L}_{\rm baryon}^{ {\cal M}_q } = {\rm Tr} \left( \left( \xi {\cal M}_q \xi 
+ \xi^\dagger {\cal M}_q^\dagger \xi^\dagger \right) {\lambda^a \over 2} \right) {\cal H}^a
+{1 \over N_c}{\rm Tr} \left( \left( \xi {\cal M}_q \xi 
+ \xi^\dagger {\cal M}_q^\dagger \xi^\dagger \right) {I \over \sqrt{6}} \right) {\cal H}^0,
\end{eqnarray}
where $a=3,8,9$.
The baryon $1/N_c$ expansion of the QCD 1-body quark operator $(\bar q \lambda^a q)/2$ is given by
\begin{eqnarray}
{\cal H}^a = \sum_{n=1}^{N_c} b_n {1 \over {N_c^{n-1}}} {\cal D}_n^a,
\end{eqnarray}
where ${\cal D}_1^a = T^a$, ${\cal D}_2^a =\left\{ J^i, G^{ia}\right\}$, and 
${\cal D}_{n+2}^a =\left\{ J^2, {\cal D}_n^a\right\}$.  The baryon $1/N_c$ expansion 
of this scalar density reduces to
\begin{eqnarray}
{\cal H}^a = b_1 T^a + b_2 {1 \over N_c} \left\{ J^i, G^{ia}\right\} + b_3 {1 \over N_c^2}\left\{ J^2, 
T^a\right\} ,
\end{eqnarray}
for $N_c =3$.
Nonet flavor symmetry of the first term in Eq.~(\ref{lagmq}) is broken by
the second term which gives a nonet symmetry-breaking contribution to the singlet $m_q$-dependent baryon
scalar density ${\cal H}^9 \equiv {\cal H}^0$ at relative order $1/N_c$.

Chiral perturbation theory for baryons in the $1/N_c$ expansion uses the vertices and 
propagators arising from the entire $1/N_c$ baryon chiral Lagrangian.
In an arbitrary Lorentz frame, the baryon propagator is given by 
$i/(k\cdot v - \Delta)$ where
\begin{eqnarray}
\Delta = M_{I} - M_{\rm ext}
\end{eqnarray}
is the mass difference of the $I$ = intermediate or internal baryon and the external baryon.
In heavy baryon chiral perturbation theory, Feynman diagrams are computed for soft pions and off-shell
baryons with momenta $k \sim {\cal O}(1)$.  
Pion exchange only couples baryons with spins differing by order unity. 
For baryons at the bottom of the spin tower with spin $J \sim {\cal O}(1)$, 
the baryon hyperfine mass splitting is ${\cal O}(1/N_c)$,
while for baryons at the top of the spin tower with $J \sim {\cal O}(N_c)$, 
the baryon hyperfine mass splitting is ${\cal O}(1)$.
The baryon mass splittings from terms dependent on quark masses are ${\cal O}(1)$, since the change in
baryon flavor quantum numbers is ${\cal O}(1)$ for pion exchange.
Thus, the baryon propagator can be expanded in a binomial expansion in baryon mass splittings
\begin{eqnarray}
\left({i \over {k \cdot v - \Delta} }\right)
=\left({i \over {k \cdot v} }\right) \sum_{n=0}^{\infty} \left( {\Delta \over {k \cdot v}} \right)^n\ .
\end{eqnarray}
In the
rest frame of the baryon, this reduces to
\begin{eqnarray}\label{bprop}
\left({i \over {k^0 - \Delta } }\right)
=\left({i \over {k^0 } }\right) \sum_{n=0}^{\infty} \left( {\Delta \over k^0} \right)^n\, .
\end{eqnarray}
The binomial expansion of the baryon propagator is valid when
the baryon mass difference is treated as
a $c$-number.  We will use this expansion in the next section to   
derive large-$N_c$ consistency conditions for the baryon mass operator ${\cal M}$.

\section{Baryon-Pion Scattering Amplitudes}

The amplitude for a baryon and pion to scatter to a final state consisting of a single baryon
and $(n-1)$ pions is ${\cal O}(N_c^{1- n/2})$ by large-$N_c$ power counting~\cite{witten}.  
(For recent 
reviews of large-$N_c$ power counting, consult Refs.~\cite{jannrev,mleshouches}).  The 
scattering amplitude is given by
\begin{eqnarray}
{\cal A} = {\cal A}_{\rm vertex} + {\cal A}_{\rm tree} + {\cal A}_{\rm loop},
\end{eqnarray}
where 
${\cal A}_{\rm vertex}$ refers to the amplitude produced by contact $n$-meson--baryon-baryon 
vertex graphs; 
${\cal A}_{\rm tree}$ denotes the amplitude obtained from all other tree diagrams;
and ${\cal A}_{\rm loop}$ represents the amplitude obtained from all loop diagrams.  
Each of these terms is at most ${\cal O}(N_c^{1-n/2})$.
Each vertex diagram is individually ${\cal O}(N_c^{1- n/2})$, so ${\cal A}_{\rm vertex}$ is leading
order.  
${\cal A}_{\rm tree}$ also is  
${\cal O}(N_c^{1- n/2})$, but individual tree diagrams may grow with higher powers of $N_c$.  
For example,
a tree diagram with $n$ separate
1-pion-baryon-baryon vertices yields an amplitude which is ${\cal O}(N_c^{n/2})$, since each
pion-baryon-baryon vertex is ${\cal O}(\sqrt{N_c})$.
Only the sum of all tree diagrams with $n$ pion--baryon-baryon vertices
is ${\cal O}(N_c^{1- n/2})$.  Thus, the individual amplitudes for
tree diagrams with $n$ pion--baryon-baryon vertices must cancel exactly to $(n-1)$ powers of
$N_c$~\cite{lamliu}.  These exact cancellations must
result from the operator structure of the tree amplitude.  
The loop amplitude
${\cal A}_{\rm loop}$ is equal to
\begin{eqnarray}
{\cal A}_{\rm loop}= \sum_{L=1}{\cal A}_{\rm loop}^{(L)},
\end{eqnarray}
where ${\cal A}_{\rm loop}^{(L)}$ denotes the scattering amplitude 
obtained from all diagrams containing $L$ loops, $L \ge 1$.  
The amplitude
${\cal A}_{\rm loop}^{(L)}$ is suppressed by a relative factor of $1/N_c^L$ compared to the 
leading in $N_c$ amplitude, and so is order $N_c^{1 -n/2 -L}$. 
Thus, the leading in $N_c$ portion of the scattering amplitude is equal to the amplitude of the
contact vertices and the leading in $N_c$ portion of the tree amplitude ${\cal A}_{\rm tree}$.
It is important to realize, however, that individual $L$-loop diagrams can grow with higher powers
of $N_c$, so that exact cancellations result from the operator structure of
the loop amplitude.  Loop cancellations and the effect of baryon mass
splittings on loop corrections are discussed in Ref.~\cite{fhjm}, which considers the
renormalization of the baryon axial vector couplings at one-loop.
We address the issue of tree diagram cancellations to all orders in
baryon mass splittings in the remainder of this paper. 

\subsection{$B + \pi \rightarrow B^\prime + \pi$}

We begin by analyzing the simplest baryon-pion scattering process 
$B + \pi \rightarrow B^\prime + \pi$ in large $N_c$, where $\pi$ denotes one of the nine 
pseudo-Goldstone
mesons $\pi$, $K$, $\eta$ and $\eta^\prime$.  The scattering process is considered for soft pions with
energies of order unity. 
The tree amplitude for this process is computed from the two tree diagrams displayed in
Fig.~1~\cite{dm93}.  In the baryon rest frame,  
\begin{eqnarray}\label{amptreepp}
{\cal A}_{\rm tree} \left(B + \pi \rightarrow B^\prime + \pi\right)= 
-{1 \over f^2}{\bf k}^i {\bf k}^{\prime j} \times
\left({{A^{jb} A^{ia}} \over {\left( k^0 - M_I + M \right)}}
+ {{A^{ia} A^{jb}} \over {\left( -k^0 - M_I + M^\prime \right)}} \right),
\end{eqnarray}
where $k^0$ is the energy of the incoming pion, $M$ is the mass of the 
initial baryon $B$, $M^\prime$ is the mass of the final baryon $B^\prime$, $M_I$ is the mass of the 
intermediate or internal baryon propagating in the diagram, and the energy of the 
outgoing pion is $k^0 + M - M^\prime$ by energy conservation.  The incoming and outgoing pions have
spin-flavor labels $ia$ and $jb$, respectively, and couple to the baryon axial vector currents $A^{ia}$
and $A^{jb}$.  
Each of the two tree diagrams in Fig.~1 
contributes to the amplitude at ${\cal O}(N_c)$ since each pion--baryon-baryon vertex is 
${\cal O}(\sqrt{N_c})$, but the total amplitude is at most ${\cal O}(1)$ by large-$N_c$ power
counting rules.  Thus, the leading ${\cal O}(N_c)$ amplitudes of the two diagrams must cancel exactly.
This cancellation requirement results in large-$N_c$ consistency conditions.
Large-$N_c$ consistency conditions for the scattering amplitude are derived from 
Eq.~(\ref{amptreepp}) by expanding each baryon 
propagator in a power series in baryon mass differences over the pion energy, 
\begin{eqnarray}\label{pptree}
{\cal A}_{\rm tree} \left(B + \pi \rightarrow B^\prime + \pi\right)&&=
-{1 \over f^2}{\bf k}^i {\bf k}^{\prime j} \times \left( \vphantom{{(M-M_I)} \over (k^0)^2}
{1 \over k^0} \left[A^{jb},A^{ia}\right] \right. \nonumber\\
&&-{{(M-M_I)} \over (k^0)^2}A^{jb}A^{ia}
-{{(M^\prime-M_I)} \over (k^0)^2}A^{ia}A^{jb} \\
&&+{{(M-M_I)^2} \over (k^0)^3}A^{jb}A^{ia}
\left.-{{(M^\prime-M_I)^2} \over (k^0)^3}A^{ia}A^{jb} + \cdots \right), \nonumber
\end{eqnarray}
where the ellipsis refers to terms proportional to higher powers of baryon mass differences.  
The terms in Eq.~(\ref{pptree}) are
proportional to $c$-number baryon mass differences and can be rewritten in terms of the baryon mass 
operator ${\cal M}$ as
\begin{eqnarray}
{\cal A}_{\rm tree} \left(B + \pi \rightarrow B^\prime + \pi\right)&&=
-{1 \over f^2}{\bf k}^i {\bf k}^{\prime j} \times \left(
{1 \over k^0} \left[A^{jb},A^{ia}\right]
+{1 \over (k^0)^2} \left[A^{jb},\left[{\cal M},A^{ia}\right]\right] \right. \\
&&+\left.{1 \over (k^0)^3} \left[A^{jb},\left[{\cal M},\left[{\cal M},A^{ia}\right]\right]\right]
+ \cdots \right), \nonumber
\end{eqnarray}
or
\begin{eqnarray}\label{i}
{\cal A}_{\rm tree} \left(B + \pi \rightarrow B^\prime + \pi\right)=
-{1 \over f^2}{\bf k}^i {\bf k}^{\prime j} \times \left(
{1 \over {(k^0)}}\sum_{n=0}^{\infty} {1 \over {(k^0)^n}} 
\left[A^{jb},\right.\underbrace{\left[{\cal M},\left[{\cal M}, \cdots 
\left[{\cal M}, \vphantom{A^{ia}}\right.\right.\right.}_{ n \ \rm insertions}
A^{ia}\underbrace{\left.\left.\left.\vphantom{A^{ia}}\right]\cdots\right]\right]}
\left.\vphantom{A^{ia}}\right] \right),
\end{eqnarray}
where $n$ refers to the number of insertions of the 
baryon mass operator ${\cal M}$ commuted with $A^{ia}$.
Notice that each insertion of the baryon mass operator
${\cal M}$ is accompanied by a commutator, so that 
the leading ${\cal O}(N_c)$ singlet piece of the baryon mass cancels out of the expression exactly 
and only the 
residual baryon mass operator gives a nonvanishing contribution.
(For the moment, we neglect
$SU(3)$ flavor symmetry-breaking, so the residual baryon mass operator is equal to the
hyperfine mass operator ${\cal M}_{\rm hyperfine}$.  When $SU(3)$ flavor breaking is not neglected,
the residual baryon mass operator also includes baryon mass splittings due to quark masses.) 

The large-$N_c$ consistency conditions for $B + \pi \rightarrow B^\prime + \pi$ scattering follow
directly from Eq.~(\ref{i}) for $k^0 \sim {\cal O}(1)$.  Each term with a different kinematic 
dependence on $k^0$ must individually
satisfy the large-$N_c$ power counting rule. 
The constraint that the tree amplitude be at most ${\cal O}(1)$ yields the large-$N_c$ consistency 
conditions
\begin{eqnarray}\label{bppcc}
\left[A^{jb},A^{ia}\right] &\ltap& {\cal O}\left(N_c\right), \nonumber\\
\left[A^{jb},\left[{\cal M},A^{ia}\right]\right]&\ltap& {\cal O}\left(N_c\right), \nonumber\\
\left[A^{jb},\left[{\cal M},\left[{\cal M},A^{ia}\right]\right]\right]
&\ltap& {\cal O}\left(N_c\right), \\
&\vdots& \nonumber
\end{eqnarray}
since the factor of $1/f^2$ in the expression for the amplitude contains an implicit factor of $1/N_c$.
The large-$N_c$ consistency conditions in Eq.~(\ref{bppcc})
can be written more compactly as
\begin{eqnarray}
\left[A^{jb},\right.\underbrace{\left[{\cal M},\left[{\cal M}, \cdots 
\left[{\cal M}, \vphantom{A^{ia}}\right.\right.\right.}_{ n \ \rm insertions}
A^{ia}\underbrace{\left.\left.\left.\vphantom{A^{ia}}\right]\cdots\right]\right]}
\left.\vphantom{A^{ia}}\right]
\ltap {\cal O}\left(N_c\right),
\end{eqnarray}
for all $n$ starting with $n=0$.

If we restrict our attention to baryons with spins $J \sim {\cal O}(1)$,
only a few of the above inequalities can be saturated and contribute to the
scattering amplitude at leading order.  This simplication occurs because 
operators with more powers of $J$ can be neglected relative to the operators with fewer 
powers of $J$ for baryons with $J \sim {\cal O}(1)$.
For baryons with spin $J \sim {\cal O}(1)$,
only the commutators $\left[A^{jb},A^{ia}\right]$ and$\left[A^{jb},\left[{\cal M},A^{ia}\right]\right]$ 
contain an ${\cal O}(N_c)$ piece.  The leading ${\cal O}(N_c)$ portion of these commutators is given
explicitly by 
\begin{eqnarray}
\left[A^{jb},A^{ia}\right] &=& a_1^2 \left[G^{jb},G^{ia}\right] + a_1 b_2
\left( \left[ G^{jb}, {1 \over N_c} J^iT^a\right] + \left[ {1 \over N_c} J^j T^b, G^{ia}\right] \right)
+ \cdots
\nonumber\\
&=& -i a_1^2 \left({1 \over 2}\epsilon^{ijk}d^{abc}G^{kc} + {1 \over 4}\delta^{ij}f^{abc} T^c \right)
-i {1 \over N_c} a_1 b_2 \epsilon^{ijk} \left( G^{kb} T^a + G^{ka} T^b \right) + \cdots ,\nonumber\\ 
\left[A^{jb},\left[{\cal M},A^{ia}\right]\right] &=& a_1^2 m_2 
\left[G^{jb},\left[{1 \over N_c} J^2,G^{ia}\right]\right]+ \cdots \\
&=& a_1^2 m_2 {1 \over N_c}
\left( \delta^{ij} \left\{ G^{\ell b},G^{\ell a} \right\} - \left\{ G^{ib}, G^{ja} \right\} + {1
\over 2} \epsilon^{ijk}f^{abc} J^kT^c \right) + \cdots, \nonumber
\end{eqnarray}
where the ellipses denote terms which are subleading in $1/N_c$ compared to terms 
which have been retained.
Thus, the 
leading ${\cal O}(1)$ portion of the tree amplitude for
$B + \pi \rightarrow B^\prime + \pi$ scattering is
\begin{eqnarray}\label{lpptree}
{\cal A}_{\rm tree} &&\left(B + \pi \rightarrow B^\prime + \pi\right)=
-{1 \over f^2}{\bf k}^i {\bf k}^{\prime j} \times \left(\vphantom{{1 \over {(k^0)}}}\right. \nonumber\\
&&{1 \over {(k^0)}}\left(-i a_1^2 \left({1 \over 2}\epsilon^{ijk}d^{abc}G^{kc} 
+ {1 \over 4}\delta^{ij}f^{abc} T^c \right)
-i {1 \over N_c} a_1 b_2 \epsilon^{ijk} \left( G^{kb} T^a + G^{ka} T^b \right) \right)\\
&&+ \left.{1 \over {(k^0)^2}}a_1^2 m_2 {1 \over N_c}
\left( \delta^{ij} \left\{ G^{\ell b},G^{\ell a} \right\} - \left\{ G^{ib}, G^{ja} \right\} + {1
\over 2} \epsilon^{ijk}f^{abc} J^kT^c \right) \right)  \nonumber
\end{eqnarray}
in the $SU(3)$ flavor symmetry limit.

The vertex amplitude contributes an additional ${\cal O}(1)$ piece from the 2-pion--baryon-baryon
contact interaction shown in Fig.~2,
\begin{eqnarray}\label{lppvertex}
{\cal A}_{\rm vertex}\left(B + \pi \rightarrow B^\prime + \pi\right) &=& -{1 \over {2f^2}}
\left( 2 k^0 + M - M^\prime \right) f^{abc} T^c \, .
\end{eqnarray}

Thus,
the leading in ${\cal O}(1)$
amplitude is given by the sum of Eqs.~(\ref{lpptree}) and (\ref{lppvertex})
for baryons with spins $J \sim {\cal O}(1)$ in the limit of $SU(3)$ flavor symmetry.

Notice that for two flavors of light quark flavors $N_F=2$ when (i) there is no $d$-symbol; (ii) the
$f$-symbol reduces to the $\epsilon$-symbol; and (iii) the flavor generator $T^a$ reduces to the isospin
generator $I^a$; the leading ${\cal O}(1)$ portion of the tree amplitude for
$B + \pi \rightarrow B^\prime + \pi$ scattering for baryons with $J \sim I \sim {\cal O}(1)$ reduces
to
\begin{eqnarray}\label{35}
{\cal A}_{\rm tree} \left(B + \pi \rightarrow B^\prime + \pi\right)=
-{1 \over f^2}{\bf k}^i {\bf k}^{\prime j} \times \left(
{1 \over {(k^0)^2}}a_1^2 m_2 {1 \over N_c}
\left( \delta^{ij} \left\{ G^{\ell b},G^{\ell a} \right\} - \left\{ G^{ib}, G^{ja} \right\}
\right) \right) ,
\end{eqnarray}
which originates from the commutator 
$\left[A^{jb},\left[{\cal M},A^{ia}\right]\right]$ that is first order in the baryon mass operator.
The commutator $\left[A^{jb},A^{ia}\right]$ that is zeroth order in the baryon mass operator yields
a subdominant contribution to the tree amplitude, since the commutator of two axial vector baryon
currents vanishes to two powers of $N_c$ for $SU(4)$ spin-flavor symmetry~\cite{dm93}.  The vertex
contribution also is of subleading order for $N_F=2$, so the leading ${\cal O}(1)$ scattering
amplitude is equal to the tree amplitude in Eq.~(\ref{35}).

When explicit $SU(3)$ symmetry breaking is included, there are additional contributions 
to the scattering amplitude.  For example, 2-pion--baryon-baryon vertices arising from 
terms in the $1/N_c$ baryon chiral Lagrangian
terms with insertions of the quark mass matrix contribute to the vertex scattering amplitude for
$B+ \pi \rightarrow B^\prime + \pi$ scattering at zero energy.  $SU(3)$
flavor symmetry breaking also appears in the tree amplitude through the baryon mass operator ${\cal
M}$, which now contains flavor-dependent mass splittings.  We will not evaluate these contributions
explicitly.

\subsection{$B + \pi \rightarrow B^\prime + \pi + \pi$}

The scattering process $B + \pi \rightarrow B^\prime + \pi + \pi$ can be analyzed in a similar manner.

The scattering amplitude ${\cal A}\left(B + \pi \rightarrow B^\prime + \pi + \pi\right)$ is 
${\cal O}(1 / \sqrt{N_c} )$ by large-$N_c$ power counting rules.  
The 3-meson--baryon-baryon vertex contribution to the scattering amplitude is explicitly 
${\cal O}(1 / \sqrt{N_c} )$, and involves no subtle cancellations.  The tree-diagram
contribution arises from the six diagrams containing three 1-meson--baryon-baryon vertices 
displayed in Fig.~3 and the six diagrams containing one 1-meson--baryon-baryon vertex and one
2-meson--baryon-baryon vertex shown in Fig.~4.  We consider these two sets of diagrams separately
because the cancellations are disjoint.

The tree amplitude from the diagrams in Fig.~3 is given by       
\begin{eqnarray}
{\cal A}^{\rm Fig.\ 3}_{\rm tree} &&\left(B + \pi \rightarrow B^\prime + \pi + \pi \right)=
{i \over f^3}{\bf k}_\alpha^i {\bf k}_\beta^j {\bf k}_\gamma^k \times
\left( \phantom{{{ A^{kc} A^{jb} A^{ia}} \over {\left( k^0_\alpha - M_1 + M \right)}}}\right. 
\nonumber\\
&&\phantom{+}{{ A^{kc} A^{jb} A^{ia}} \over 
{\left( k^0_\alpha - M_1 + M \right)\left( k^0_\gamma - M_2 + M^\prime \right)}}
+{{ A^{jb} A^{kc} A^{ia}} \over 
{\left( k^0_\alpha - M_1 + M \right)\left( k^0_\beta - M_2 + M^\prime \right)}}\nonumber\\
&&+{{A^{kc} A^{ia} A^{jb} } \over 
{\left( -k^0_\beta - M_1 + M \right)\left( k^0_\gamma - M_2 + M^\prime \right)}}
+{{A^{ia} A^{kc} A^{jb} } \over 
{\left( -k^0_\alpha - M_2 + M^\prime \right)\left( -k^0_\beta - M_1 + M \right)}} \\
&&+{{A^{ia} A^{jb} A^{kc} } \over 
{\left( -k^0_\alpha - M_2 + M^\prime \right)\left( -k^0_\gamma - M_1 + M \right)}}
+\left.{{A^{jb} A^{ia} A^{kc} } \over 
{\left( k^0_\beta - M_2 + M^\prime \right)\left( -k^0_\gamma - M_1 + M \right)}} \right),\nonumber
\end{eqnarray}
where
$k^0_\alpha$ denotes the energy of the incoming pion with spin-flavor labels $ia$,
$k^0_\beta$ denotes the energy of the outgoing pion with spin-flavor labels $jb$, and
$k^0_\gamma$ denotes the energy of the outgoing pion with spin-flavor labels $kc$.  
Each of the pion energies is taken to be ${\cal O}(1)$.
The
masses $M$ and $M^\prime$ are the masses of the initial baryon $B$ and the final baryon $B^\prime$,
respectively, while $M_1$ and $M_2$ are the masses of the first and second intermediate baryons,
respectively.  Energy conservation implies that $k^0_\alpha + M = k^0_\beta + k^0_\gamma + M^\prime$,
so the energy $k^0_\alpha$ can be eliminated from the expression using this relation.  
Each of the six diagrams produces a contribution of ${\cal O}\left({N_c}^{3/2}\right)$ to the amplitude,
but the overall amplitude is only ${\cal O}\left(1/\sqrt{N_c}\right)$, so exact cancellations 
must occur between the diagrams to two powers of $N_c$.
Expanding the baryon
propagators in powers of baryon mass differences over pion energies, and rewriting the expression using the
baryon mass operator ${\cal M}$ yields the lengthy expression given in Appendix~A.  Consistency of the
large-$N_c$ limit requires that the terms with different kinematic dependence must each be 
${\cal O}\left(1/\sqrt{N_c}\right)$.  Taking into account the implicit $(1/\sqrt{N_c})^3$ dependence 
of the overall factor $1/f^3$, the linear combinations of commutators appearing in the
expression must each be $\ltap {\cal O}(N_c)$.
Thus, the large-$N_c$ consistency conditions obtained 
from the tree diagrams in Fig.~3 are:
\begin{eqnarray}\label{ampzero}
\left[ A^{kc}, \left[ A^{ia},  A^{jb} \right] \right] \ltap {\cal O}\left( N_c\right),
\end{eqnarray}
to zeroth order in baryon mass differences,
\begin{eqnarray}\label{ampone}
&&\left[ A^{kc}, \left[ A^{ia}, \left[ {\cal M}, A^{jb} \right] \right] \right]
\ltap {\cal O}\left( N_c\right), \nonumber\\
&&\left[\left[{\cal M}, A^{kc} \right], \left[A^{ia}, A^{jb} \right] \right]
\ltap {\cal O}\left( N_c\right),
\end{eqnarray}
to first order in baryon mass differences, and
\begin{eqnarray}\label{amptwo}
&&\left[\left[{\cal M},\left[{\cal M}, A^{kc} \right]\right], \left[A^{ia}, A^{jb} \right]\right]
\ltap {\cal O}\left( N_c\right), \nonumber\\
&&\left[A^{kc}, \left[A^{ia}, \left[{\cal M}, \left[{\cal M}, A^{jb}\right]\right]\right]\right]
\ltap {\cal O}\left( N_c\right), \\
&&\left[\left[{\cal M}, A^{kc} \right],\left[ A^{ia}, \left[{\cal M}, A^{jb} \right]\right]\right]
\ltap {\cal O}\left( N_c\right), \nonumber
\end{eqnarray}
to second order in baryon mass differences.  In general, 
to $n^{\rm th}$ order in baryon mass differences, the consistency conditions restrict
all possible commutators obtained from 
$\left[ A^{kc}, \left[ A^{ia},  A^{jb} \right] \right]$ with a total of $n$ additional
commutators of ${\cal M}$ dressing the baryon axial vector currents $A^{kc}$ and $A^{jb}$
to be $\ltap {\cal O}(N_c)$.

If we restrict our attention to baryons with
spin $J \sim {\cal O}(1)$,
not all of these commutators contribute to the $B + \pi \rightarrow B^\prime + \pi + \pi$
scattering amplitude at leading order ${\cal O}(1/\sqrt{N_c})$.  
For baryons with spin $J \sim {\cal O}(1)$, only the commutators
of Eq.~(\ref{ampzero}) and Eq.~(\ref{ampone}) and the last two commutators in Eq.~(\ref{amptwo})
contain an ${\cal O}(N_c)$ piece.  (The first commutator in Eq.~(\ref{amptwo}) is necessarily
suppressed in the $1/N_c$ expansion for baryons with spin $J \sim {\cal O}(1)$ since there
are not enough commutators to get rid of all of the $J$'s coming from the baryon mass operators.) 
The leading ${\cal O}(N_c)$ portions of these commutators can be evaluated explicitly,
but are rather lengthy and will be suppressed.

There are additional cancellations involving the
2-pion--baryon-baryon vertices
occurring between the six tree diagrams displayed in Fig.~4.
The amplitude produced by the diagrams in Fig.~4 is given by
\begin{eqnarray}
{\cal A}^{\rm Fig.\ 4}_{\rm tree} &&\left(B + \pi \rightarrow B^\prime + \pi + \pi \right)=
-{i \over {2f^3}}\times \left(
\vphantom{{k_\alpha^0 \over k_\alpha^0}} \right.\nonumber\\
&&\phantom{-}
{\bf k}_\alpha^i \left(k_\gamma^0 - k_\beta^0 \right) 
f^{dbc} \left( { {T^d A^{ia}}  \over {(k_\alpha^0 + M - M_I)}}
+{ {A^{ia} T^d}  \over {(-k_\alpha^0 + M^\prime - M_I)}} \right) \nonumber\\
&&-{\bf k}_\gamma^i \left(k_\alpha^0 + k_\beta^0 \right)
f^{dab} \left( { {T^d A^{ic}} \over {(-k_\gamma^0 + M - M_I)}}
+{{A^{ic} T^d} \over {(k_\gamma^0 + M^\prime - M_I)}} \right) \\
&&-{\bf k}_\beta^i \left(k_\alpha^0 + k_\gamma^0 \right)
f^{dac} \left( {{T^d A^{ib}} \over {(-k_\beta^0 + M - M_I)}}
+{{A^{ib} T^d}  \over {(k_\beta^0 + M^\prime - M_I)}} \right)
\left. \vphantom{{k_\alpha^0 \over k_\alpha^0}}\right),\nonumber
\end{eqnarray}
where $M_I$ is the mass of the intermediate baryon.
The 2-pion--baryon-baryon vertex is ${\cal O}(1)$ whereas the 1-pion--baryon-baryon
vertex is ${\cal O}(\sqrt{N_c})$, so each of the six diagrams is ${\cal O}(\sqrt{N_c})$.  The large
$N_c$ power counting rule that the $B + \pi \rightarrow B^\prime + \pi + \pi$
scattering amplitude is ${\cal O}(1/\sqrt{N_c})$ implies that
the diagrams cancel exactly to one power in $N_c$. 
Expanding the terms in baryon mass differences over pion energies, and rewriting the expression in
terms of the baryon mass operator ${\cal M}$ yields
\begin{eqnarray}\label{vppap}
{\cal A}^{\rm Fig.\ 4}_{\rm tree} 
&&\left(B + \pi \rightarrow B^\prime + \pi + \pi \right)=-{i \over {2f^3}}\times \left(
\vphantom{{k_\alpha^0 \over k_\alpha^0}} \right.\nonumber\\
&&
{\bf k}_\alpha^i {{\left(k_\gamma^0 - k_\beta^0 \right)} \over k_\alpha^0 }
f^{dbc} \left( \sum_{n=0}^\infty \left({1 \over k_\alpha^0}
\right)^{n}\left[ T^d , \left[{\cal M},\left[{\cal M}, \cdots \left[ {\cal M},
A^{ia}\right]\cdots \right]\right]\right] \right)\nonumber\\ 
&&+{\bf k}_\gamma^i {{\left(k_\alpha^0 + k_\beta^0 \right)} \over k_\gamma^0 }
f^{dab} \left( \sum_{n=0}^\infty \left({{-1} \over k_\gamma^0}
\right)^{n}\left[T^d , \left[{\cal M},\left[{\cal M}, \cdots \left[ {\cal M},
A^{ic}\right]\cdots \right]\right]\right] \right) \\ 
&&+{\bf k}_\beta^i {{\left(k_\alpha^0 + k_\gamma^0 \right)} \over k_\beta^0 }
f^{dac} \left( \sum_{n=0}^\infty \left({{-1} \over k_\beta^0}
\right)^{n}\left[ T^d , \right. 
\underbrace{\left[{\cal M},\left[{\cal M}, \cdots \left[ {\cal M},
\vphantom{A^{ib}}\right.\right.\right.}_{ n \ \rm insertions}
A^{ib}
\underbrace{\left.\left.\left.\vphantom{A^{ib}}\right]\cdots \right]\right]}
\left.\vphantom{A^{ib}}\right] \right)  
\left. \vphantom{{k_\alpha^0 \over k_\alpha^0}}\right),\nonumber
\end{eqnarray}
where the terms in the summation have $n$ commutators of ${\cal M}$ with the baryon axial current.

The large-$N_c$ consistency conditions for the diagrams in Fig.~4 
are derived from Eq.~(\ref{vppap}) for
pion energies of order unity.  The constraint that the tree amplitude be at most 
${\cal O}(1/\sqrt{N_c})$ yields the large-$N_c$ consistency conditions
\begin{eqnarray}\label{tbppcc}
f^{dbc} \left[T^{d},A^{ia}\right] &\ltap& {\cal O}\left(N_c\right), \nonumber\\
f^{dbc} \left[T^{d},\left[{\cal M},A^{ia}\right]\right]&\ltap& {\cal O}\left(N_c\right), \nonumber\\
f^{dbc} \left[T^{d},\left[{\cal M},\left[{\cal M},A^{ia}\right]\right]\right]
&\ltap& {\cal O}\left(N_c\right), \\
&\vdots& \nonumber
\end{eqnarray}
or
\begin{eqnarray}\label{tconstraint}
f^{dbc} \left[T^{d},\right.\underbrace{\left[{\cal M},\left[{\cal M}, \cdots 
\left[{\cal M}, \vphantom{A^{ia}}\right.\right.\right.}_{ n \ \rm insertions}
A^{ia}\underbrace{\left.\left.\left.\vphantom{A^{ia}}\right]\cdots\right]\right]}
\left.\vphantom{A^{ia}}\right]
\ltap {\cal O}\left(N_c\right),
\end{eqnarray}
for all $n$ starting with $n=0$.

For baryons with spin $J \sim {\cal O}(1)$, only the first commutator in Eq.~(\ref{tbppcc})
can contribute to the scattering amplitude at leading order.  Thus, the leading 
${\cal O}(1/\sqrt{N_c})$
portion of the amplitude from the diagrams in Fig.~4 is given by
\begin{eqnarray}\label{amp21}
&&{\cal A}^{\rm Fig.\ 4}_{\rm tree} \left(B + \pi \rightarrow B^\prime + \pi + \pi \right)
\nonumber\\
&&=-{i \over {2f^3}}\times
\left( {\bf k}_\alpha^i {{\left(k_\gamma^0 - k_\beta^0 \right)} \over k_\alpha^0 }
f^{dbc} \left[T^d, A^{ia}\right] 
 -{\bf k}_\gamma^i {{\left(k_\alpha^0 + k_\beta^0 \right)} \over k_\gamma^0 }
f^{dab} \left[T^d, A^{ic}\right] 
-{\bf k}_\beta^i {{\left(k_\alpha^0 + k_\gamma^0 \right)} \over k_\beta^0 }
f^{dac} \left[T^d, A^{ib}\right]  \right), \\ 
&&={1 \over {2f^3}}\times \left(
{\bf k}_\alpha^i {{\left(k_\gamma^0 - k_\beta^0 \right)} \over k_\alpha^0 } f^{dbc} f^{dag}
-{\bf k}_\gamma^i {{\left(k_\alpha^0 + k_\beta^0 \right)} \over k_\gamma^0 } f^{dab} f^{dcg}
-{\bf k}_\beta^i {{\left(k_\alpha^0 + k_\gamma^0 \right)} \over k_\beta^0 }
f^{dac} f^{dbg} \right)A^{ig}, \nonumber
\end{eqnarray}
for baryons with spin $J \sim {\cal O}(1)$,
where the second equality follows because the baryon axial vector current transforms as a 
flavor adjoint under $SU(3)$ flavor symmetry.  The amplitude Eq.~(\ref{amp21})  
is now manifestly $\ltap {\cal O}\left(1/\sqrt{N_c}\right)$ since the matrix elements of the baryon axial
vector current are $\ltap {\cal O}(N_c)$ and the overall factor of $1/f^3$ contains an implicit factor
of $(1/\sqrt{N_c})^3$.

The vertex amplitude contributes an ${\cal O}(1/\sqrt{N_c})$ piece from the contact
3-pion--baryon-baryon interaction given by
\begin{eqnarray}\label{vertexppp}
{\cal A}_{\rm vertex} \left(B + \pi \rightarrow B^\prime + \pi + \pi \right) &=&
{i \over {3! f^3}}
\left( 
-{\bf k}_\alpha^i \left(f^{bed} f^{cae} +f^{ced} f^{bae} \right)
+ {\bf k}_\beta^i \left(f^{ced} f^{abe} +f^{aed} f^{cbe} \right) \right. \nonumber\\
&&\qquad\qquad
\left.+{\bf k}_\gamma^i \left(f^{aed} f^{bce} +f^{bed} f^{ace} \right)
\right)
A^{id} \, .
\end{eqnarray}

Thus, the leading ${\cal O}(1/\sqrt{N_c})$ portion of the tree amplitude for 
$B + \pi \rightarrow B^\prime + \pi + \pi$ scattering is given in the $SU(3)$ flavor symmetry
limit by the sum of 
the leading ${\cal O}(1/\sqrt{N_c})$ part of the amplitude from the diagrams in Fig.~3, and
Eqs.~(\ref{amp21}) and~(\ref{vertexppp}).

When explicit $SU(3)$ symmetry breaking is included, there are additional contributions to the 
scattering amplitude.    $SU(3)$
flavor symmetry breaking appears in the tree amplitudes through the baryon mass operator ${\cal
M}$, which now contains flavor-dependent mass splittings.  In addition, there is a contribution
to the 2-pion--baryon-baryon vertices in the diagrams in Fig.~4 from quark mass-dependent terms
in the chiral Lagrangian.  
The 2-pion--baryon-baryon vertices from the $1/N_c$
baryon chiral Lagrangian linear in the quark mass matrix given in Eq.~(\ref{lagmq}) 
are proportional to the
baryon scalar density
operator ${\cal H}^a$, $a=3,8,9$.  Large-$N_c$ consistency conditions for these
2-pion--baryon-baryon vertices follow from the constraint that the scattering amplitude 
is $\ltap {\cal O}(1/ \sqrt{N_c})$.  These additional large-$N_c$ consistency conditions are:  
\begin{eqnarray}\label{hconstraint}
f^{dbc} \left[{\cal H}^{d},\right.\underbrace{\left[{\cal M},\left[{\cal M}, \cdots 
\left[{\cal M}, \vphantom{A^{ia}}\right.\right.\right.}_{ n \ \rm insertions}
A^{ia}\underbrace{\left.\left.\left.\vphantom{A^{ia}}\right]\cdots\right]\right]}
\left.\vphantom{A^{ia}}\right]
\ltap {\cal O}\left(N_c\right),
\end{eqnarray}
for all $n$ starting with $n=0$.  Notice that
Eq.~(\ref{hconstraint}) is obtained from Eq.~(\ref{tconstraint})
by replacing $T^d$ by ${\cal H}^d$.  The singlet ${\cal O}(N_c)$ piece of ${\cal H}^a$ cancels out of the
expression exactly, so that only the $a=3$ and $a=8$ components of ${\cal H}^a$ give a nonvanishing
contribution.  Thus, the pion-nucleon sigma term does not contribute to the scattering amplitude.  
Similar large-$N_c$ consistency conditions also can be found for terms in
the $1/N_c$ baryon chiral Lagrangian with more insertions of the quark mass matrix.

\subsection{$B + \pi \rightarrow B^\prime + (n-1) \pi$}

The generalization to the scattering process $B + \pi \rightarrow B^\prime + (n-1) \pi$ for $n > 3$
is straightforward, although the expressions for the scattering amplitudes necessarily become
very lengthy.
The large-$N_c$ consistency conditions that follow from diagrams with $n$ 1-pion--baryon-baryon vertices
are that
\begin{eqnarray}\label{acoms}
\left[ A^{i_n a_n},   \cdots   \left[ A^{i_3 a_3}, \left[ A^{i_2 a_2}, 
A^{i_1 a_1} \right]\right] \cdots \right]
\ltap {\cal O}(N_c),
\end{eqnarray}  
and that all multicommutators of $n$ baryon axial vector currents with additional
commutators of ${\cal M}$ dressing the baryon axial vector currents in all possible
ways are constrained to be at most ${\cal O}(N_c)$. 
Because of the multiplicative factor of $(1/f)^{n}$ in the amplitude, these large-$N_c$ consistency
conditions imply that the tree scattering amplitude is ${\cal O}(N_c^{1- n/2})$.
Additional cancellations occur for diagrams with multimeson--baryon-baryon vertices, yielding
large-$N_c$ consistency conditions involving these vertices.  
All multipion--baryon-baryon vertices with an odd number
of pions derive in the flavor symmetry limit
from the pion axial vector current terms which are proportional to the baryon axial
vector current operator.  The large-$N_c$ consistency conditions involving only vertices with odd
numbers of pions therefore are of the form given in Eq.~(\ref{acoms})
where the number of axial vector currents is now less that $n$.  Multipion--baryon-baryon
vertices with an even number of pions derive in the flavor symmetry limit 
from the pion vector current terms which are proportional
to the baryon flavor operator.  Large-$N_c$ consistency conditions involving one
even-pion--baryon-baryon vertex and multiple
odd-pion--baryon-baryon vertices are given by 
\begin{eqnarray}
\left[ T^{b}, \left[ A^{i_{\ell} a_{\ell}},  \cdots   \left[ A^{i_2 a_2}, 
A^{i_1 a_1} \right] \cdots \right] \right]
\ltap {\cal O}(N_c).
\end{eqnarray}
All multicommutators deriving from this multicommutator with additional commutators of ${\cal M}$
dressing the baryon axial vector currents in all possible ways are constrained to be at most ${\cal
O}(N_c)$.  Diagrams with two even-pion--baryon-baryon vertices and multiple odd-pion--baryon-baryon
vertices yield the large-$N_c$ consistency conditions
\begin{eqnarray}
\left[ T^{b_2}, \left[ T^{b_1}, \left[ A^{i_{\ell} a_{\ell}},  \cdots   \left[ A^{i_2 a_2}, 
A^{i_1 a_1} \right]\cdots \right] \right] \right]
\ltap {\cal O}(N_c),
\end{eqnarray}
as well as all possible dressings of these multicommutators with commutators of ${\cal M}$ being restricted
to be at most ${\cal O}(N_c)$.  The
generalization to an arbitrary number of even-pion--baryon-baryon vertices is immediate.  Other
large-$N_c$ consistency conditions follow from terms in the $1/N_c$ baryon chiral Lagrangian
with insertions of the quark mass matrix.  Terms with no pions can be incorporated into the baryon mass
operator, and appear in the multicommutators through ${\cal M}$.  
Terms with even numbers of pions can occur in place of some or all
of the pion vector current couplings to
baryons.  The large-$N_c$ consistency conditions involving these flavor symmetry breaking vertices
are of the form
\begin{eqnarray}
\left[ T^{c_n}, \cdots 
\left[ {\cal T}^{c_1}, \left[ {\cal H}^{b_m}, \cdots \left[ {\cal H}^{b_1}, 
\left[ A^{i_{\ell} a_{\ell}},  \cdots   \left[ A^{i_2 a_2}, 
A^{i_1 a_1} \right] \cdots \right] \right]  \cdots \right] \right] \cdots \right]
\ltap {\cal O}(N_c).
\end{eqnarray}
All possible dressings of these multicommutators with arbitrary numbers of commutators of
the baryon mass operator ${\cal M}$ are constrained to be $\ltap {\cal O}(N_c)$ as well.

\section{Conclusions}

Exact cancellations occur in the tree amplitudes for baryon-pion scattering amplitudes at leading orders in
$N_c$.  As the number of pions involved in the scattering grows, the power in $N_c$ of the cancellation also
grows.  In this paper, we derived the  
large-$N_c$ consistency 
conditions for baryon-pion scattering processes to all orders in baryon mass splittings.
The baryon mass splittings contribute to the scattering amplitudes through commutators of the baryon mass
operator ${\cal M}$, so  
the ${\cal O}(N_c)$ singlet portion of the baryon mass cancels out of the amplitudes exactly. 
We showed 
that the leading in $N_c$ portion
of any baryon-pion scattering amplitude only requires the evaluation of terms to a finite order in
baryon mass splittings for baryons with spins $J \sim {\cal O}(1)$.  We explicitly computed the leading
order in $N_c$ scattering amplitude for the simplest processes $B+\pi \rightarrow B^\prime + \pi$ and 
$B+\pi \rightarrow B^\prime + \pi +\pi$.  
These cancellations were discussed previously
in Ref.~\cite{djm94}
to zeroth order in baryon mass splittings.
The contribution to $B+\pi \rightarrow B^\prime + \pi$ scattering at first subleading order in baryon
mass splittings also was derived in Ref.~\cite{djm94}, and was determined to be
the dominant portion of the tree amplitude for $\pi^{\pm,0}$ scatterings.  
Additional large-$N_c$ consistency conditions 
for baryon vector couplings, as well as for baryon couplings involving explicit
$SU(3)$ symmetry breaking due to the quark mass matrix, were derived as well.  There are many different
types of
cancellations occuring in the tree scattering amplitudes, and the significance of the
cancellations increases with the number of pions in the scattering process.

\acknowledgements

This work was supported in part by the Department of Energy under Grant No. DOE-FG03-97ER40546.
R.F.M. was supported in part by CONACYT (Mexico) under the UC-CONACYT agreement of cooperation,
CINVESTAV (Mexico), and a UCSD grant from the Alfred P. Sloan Foundation.  C.P.H. acknowledges
support from the Schweizerischer Nationalfonds and Holderbank-Stiftung.
E.J. was supported in part by the Alfred P. Sloan Foundation and by the National Young Investigator
program through Grant No. PHY-9457911 from the National Science Foundation.

\appendix

\section{}

The contribution to ${\cal A}_{\rm tree}\left(B + \pi \rightarrow B^\prime + \pi + \pi
\right)$ from the diagrams displayed in Fig.~3, 
written in terms of the baryon mass operator ${\cal M}$, is given by:
\begin{eqnarray}
{\cal A}^{\rm Fig. \ 3}_{\rm tree} &&\left(B + \pi \rightarrow B^\prime + \pi + \pi \right)= 
{i \over f^3}{\bf k}_\alpha^i {\bf k}_\beta^j {\bf k}_\gamma^k \times
\left( 
\phantom{{1 \over {k^0_\beta}}\left[ A^{ia} , A^{jb} \right]}\right. \nonumber\\
&&-{1 \over {(k^0_\beta + k^0_\gamma)k^0_\gamma}}\left[ A^{kc}, \left[ A^{ia} , A^{jb} \right] \right]
-{1 \over {(k^0_\beta + k^0_\gamma)k^0_\beta}}\left[ A^{jb}, \left[ A^{ia} , A^{kc} \right] \right]
\nonumber\\
&&+{1 \over {(k^0_\beta + k^0_\gamma)^2 k^0_\gamma}}
\left( \left[ A^{kc}, \left[ A^{ia}, \left[ {\cal M}, A^{jb} \right] \right] \right] 
+ \left[\left[{\cal M}, A^{kc} \right], \left[A^{ia}, A^{jb} \right] \right] \right)
\nonumber\\
&&+{1 \over {(k^0_\beta + k^0_\gamma)^2 k^0_\beta}}
\left( \left[ A^{jb}, \left[ A^{ia}, \left[ {\cal M}, A^{kc} \right] \right] \right] 
+ \left[\left[{\cal M}, A^{jb} \right], \left[A^{ia}, A^{kc} \right] \right] \right) \nonumber\\
&&+{1 \over {(k^0_\beta + k^0_\gamma) (k^0_\gamma)^2}}
\left[\left[{\cal M}, A^{kc} \right], \left[A^{ia}, A^{jb} \right] \right]
+{1 \over {(k^0_\beta + k^0_\gamma) (k^0_\beta)^2}}
\left[\left[{\cal M}, A^{jb} \right], \left[A^{ia}, A^{kc} \right] \right] \nonumber\\
&&-{1 \over {(k^0_\beta + k^0_\gamma)^3 k^0_\gamma}} \left( 
\left[\left[{\cal M},\left[{\cal M}, A^{kc} \right]\right], \left[A^{ia}, A^{jb} \right]\right]
+ \left[A^{kc}, \left[A^{ia}, \left[{\cal M}, \left[{\cal M}, A^{jb}\right]\right]\right]\right] 
\right. \nonumber\\ 
&&\left.\qquad\qquad\qquad\qquad\qquad\qquad
+ 2 \left[\left[{\cal M}, A^{kc} \right],\left[ A^{ia}, \left[{\cal M}, A^{jb} \right]\right]\right]
\right) \nonumber\\
&&-{1 \over {(k^0_\beta + k^0_\gamma)^3 k^0_\beta}} \left( 
\left[\left[{\cal M},\left[{\cal M}, A^{jb} \right]\right], \left[A^{ia}, A^{kc} \right]\right]
+ \left[A^{jb}, \left[A^{ia}, \left[{\cal M}, \left[{\cal M}, A^{kc}\right]\right]\right]\right]
\right. \\ 
&&\left. \qquad\qquad\qquad\qquad\qquad\qquad
+ 2 \left[\left[{\cal M}, A^{jb} \right],\left[ A^{ia}, \left[{\cal M}, A^{kc} \right]\right]\right]
\right) \nonumber\\
&&-{1 \over {(k^0_\beta + k^0_\gamma)^2 (k^0_\gamma)^2}} \left( 
\left[\left[{\cal M},\left[{\cal M}, A^{kc} \right]\right], \left[A^{ia}, A^{jb} \right]\right]
+ \left[\left[{\cal M}, A^{kc} \right],\left[ A^{ia}, \left[{\cal M}, A^{jb} \right]\right]\right]
\right) \nonumber\\
&&-{1 \over {(k^0_\beta + k^0_\gamma)^2 (k^0_\beta)^2}} \left( 
\left[\left[{\cal M},\left[{\cal M}, A^{jb} \right]\right], \left[A^{ia}, A^{kc} \right]\right]
+ \left[\left[{\cal M}, A^{jb} \right],\left[ A^{ia}, \left[{\cal M}, A^{kc} \right]\right]\right]
\right) \nonumber\\
&&-{1 \over {(k^0_\beta + k^0_\gamma)(k^0_\gamma)^3}}\left[\left[{\cal M}, \left[{\cal M}, A^{kc} \right]
\right], \left[A^{ia}, A^{jb} \right] \right] 
-{1 \over {(k^0_\beta + k^0_\gamma)(k^0_\beta)^3}}\left[\left[{\cal M}, \left[{\cal M}, A^{jb} \right]
\right], \left[A^{ia}, A^{kc} \right] \right] \nonumber\\
&&\left.+ \qquad\cdots 
\phantom{{1 \over {k^0_\beta}}\left[ A^{ia}\right]}\right) ,\nonumber
\end{eqnarray}
where the ellipsis denotes terms with more powers of the baryon mass operator ${\cal M}$.

\newpage

\begin{table}
\caption{$SU(2 N_F)$ Commutation Relations}
\bigskip
\label{tab:su2fcomm}
\centerline{\vbox{ \tabskip=0pt \offinterlineskip
\halign{
\strut\quad $ # $\quad\hfil&\strut\quad $ # $\quad \hfil\cr
\multispan2\hfil $\left[J^i,T^a\right]=0,$ \hfil \cr
\noalign{\medskip}
\left[J^i,J^j\right]=i\epsilon^{ijk} J^k,
&\left[T^a,T^b\right]=i f^{abc} T^c,\cr
\noalign{\medskip}
\left[J^i,G^{ja}\right]=i\epsilon^{ijk} G^{ka},
&\left[T^a,G^{ib}\right]=i f^{abc} G^{ic},\cr
\noalign{\medskip}
\multispan2\hfil$\left[G^{ia},G^{jb}\right] = {i\over 4}\delta^{ij} f^{abc} T^c + {i\over 
{2N_F}} \delta^{ab} \epsilon^{ijk} J^k + {i\over 2} \epsilon^{ijk} d^{abc} 
G^{kc}.$ \hfill\cr
}}}
\end{table}

\newpage

\begin{figure}\label{fig:bpptree}
\centerline{\epsfysize=2.5cm\epsffile{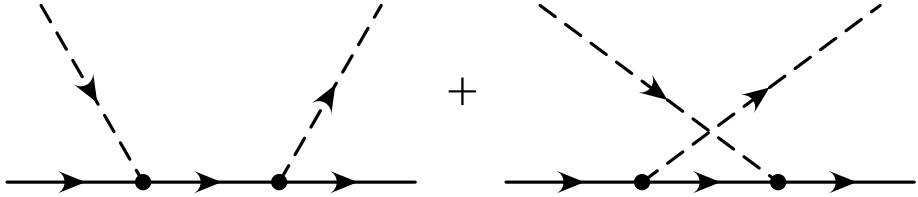}}
\centerline{}
\tighten{\caption{Tree diagrams contributing to the $B + \pi \rightarrow B^\prime + \pi$
scattering amplitude.  Each individual diagram is ${\cal O}\left(N_c\right)$,
but the sum of the two diagrams is ${\cal O}\left( 1 \right)$.}}
\end{figure}

\smallskip

\begin{figure}\label{fig:bppvertex}
\centerline{\epsfysize=2.5cm\epsffile{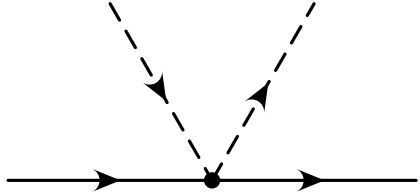}}
\centerline{}
\tighten{\caption{Vertex contribution to the scattering amplitude for 
$B + \pi \rightarrow B^\prime + \pi$.  The vertex amplitude is ${\cal O}\left(1\right)$.}}
\end{figure}

\smallskip

\begin{figure}\label{figthree}
\centerline{\epsfysize=2.5cm\epsffile{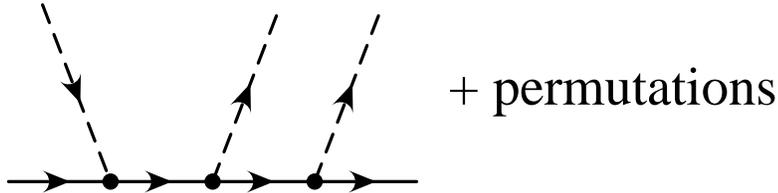}}
\centerline{}
\tighten{\caption{Tree diagrams contributing to the $B + \pi \rightarrow B^\prime + \pi + \pi$
scattering amplitude containing three ${\cal O}(\sqrt{N_c})$ pion--baryon-baryon vertices.  
Each individual diagram is ${\cal O}\left(N_c^{3/2}\right)$,
but the sum of the six diagrams is ${\cal O}\left( 1/\sqrt{N_c} \right)$.}}
\end{figure}

\smallskip

\begin{figure}\label{figfour}
\centerline{\epsfysize=2.5cm\epsffile{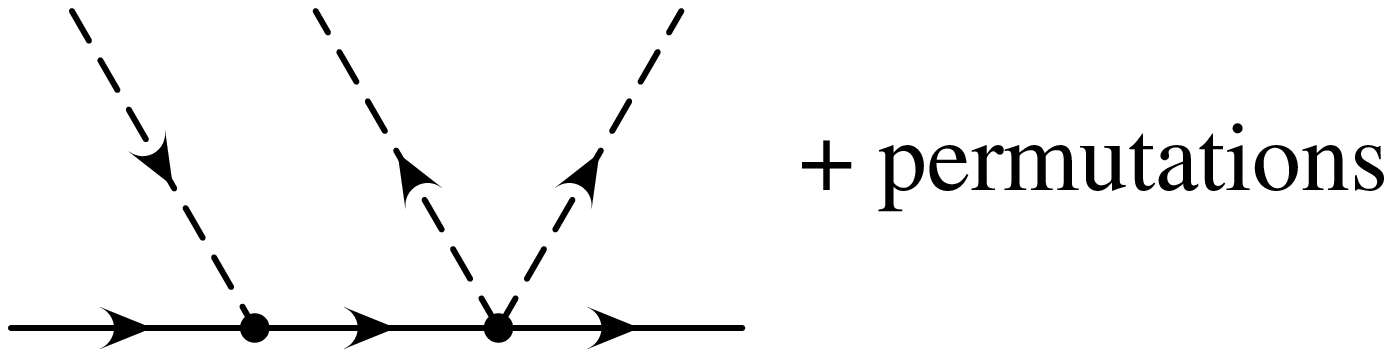}}
\centerline{}
\tighten{\caption{Tree diagrams contributing to the $B + \pi \rightarrow B^\prime + \pi + \pi$
scattering amplitude containing one ${\cal O}(\sqrt{N_c})$ pion--baryon-baryon vertex and one
${\cal O}(1)$ 2-pion--baryon-baryon vertex.  Each individual diagram is ${\cal O}\left(\sqrt{N_c}\right)$,
but the sum of the six diagrams is ${\cal O}\left( 1/\sqrt{N_c} \right)$.}}
\end{figure}


\begin{references}

\bibitem{jannrev} E. Jenkins, Annu. Rev. Nucl. Part. Sci. {\bf 48}, 81 (1998).

\bibitem{dm93} R.F. Dashen and A.V. Manohar, Phys. Lett. B{\bf 315}, 425 (1993); 438 (1993).

\bibitem{j93} E. Jenkins, Phys. Lett. B{\bf 315}, 441 (1993).

\bibitem{djm94} R.F. Dashen, E. Jenkins, and A.V. Manohar, Phys. Rev. D
{\bf 49}, 4713 (1994).

\bibitem{lamliu} C.S. Lam and K.F. Liu, Phys. Rev. Lett. {\bf 79}, 597 (1997).

\bibitem{djm95} R.F. Dashen, E. Jenkins, and A.V. Manohar, Phys. Rev. D
{\bf 51}, 3697 (1995).

\bibitem{lmr} M. Luty and J. March-Russell, Nucl. Phys. B{\bf 426}, 71 (1994).

\bibitem{cgo} C. Carone, H. Georgi and S. Osofsky, Phys. Lett. B{\bf 322}, 227 (1994).

\bibitem{thooft76} G. 't Hooft, Phys. Rev. D {\bf 14}, 3432 (1976).

\bibitem{j96} E. Jenkins, Phys. Rev. D {\bf 53}, 2625 (1996).

\bibitem{jm255} E. Jenkins and A.V. Manohar, Phys. Lett. B{\bf 255}, 558
(1991).

\bibitem{jm259} E. Jenkins and A.V. Manohar, Phys. Lett. B{\bf 259}, 353
(1991).

\bibitem{witten} E. Witten, Nucl. Phys. B {\bf 160}, 57 (1979).

\bibitem{mleshouches} A.V. Manohar, {\it Large N QCD}, Les Houches, Session LXVIII, 1997, 
Probing the Standard Model of Particle Interactions, eds. R. Gupta, A. Morel, E. de Rafael 
and F. David, (Elsevier Science, Amsterdam, 1999) 1091.

\bibitem{fhjm} R. Flores-Mendieta, C.P. Hofmann, E. Jenkins and A.V. Manohar,
UCSD/PTH 99-22.

\end{references}
\end{document}